\documentclass[12pt]{article}

\usepackage{amsfonts}
\usepackage{amsmath}
\usepackage{hyperref}
\usepackage{cite}
\usepackage{epsfig}
\usepackage{latexsym}
\usepackage{paralist}
\usepackage{fancyhdr}
\usepackage{tikz}
\usetikzlibrary{decorations.pathmorphing,patterns,calc,snakes,arrows}

\usepackage{graphicx}
\usepackage{xcolor}
\numberwithin{equation}{section}
\usepackage[vcentermath]{youngtab}
\usepackage{etex}
\usepackage{braket}
\usepackage{float}

\def\spa#1{\phantom{\fbox{\rule[-#1cm]{0cm}{0cm}}}}

\def\be{\begin{equation}}
\def\ee{\end{equation}}
\def\bea{\begin{eqnarray}}
\def\eea{\end{eqnarray}}

\def\half{{1\over 2}}

\def\del{\partial}
\def\nn{\nonumber}

\renewcommand{\thefootnote}{\fnsymbol{footnote}}


\def\cO{{\cal O}}

\begin{document}

\hfuzz=100pt
\title{{\Large \bf{Random Boundary Geometry\\ and \\Gravity Dual of $T\bar{T}$ Deformation}}}
\date{}
\author{ Shinji Hirano$^{a, c}$\footnote{
	e-mail:
	\href{mailto:shinji.hirano@wits.ac.za}{shinji.hirano@wits.ac.za}}
  \,and Masaki Shigemori$^{b, c}$\footnote{
	e-mail:
	\href{mailto:masaki.shigemori@nagoya-u.jp}{masaki.shigemori@nagoya-u.jp}}
}
\date{}

\maketitle

\thispagestyle{fancy}
\rhead{YITP-20-26}
\cfoot{}
\renewcommand{\headrulewidth}{0.0pt}

\vspace*{-1cm}
\begin{center}
$^{a}${{\it School of Physics and Mandelstam Institute for Theoretical Physics }}
\\ {{\it University of the Witwatersrand}}
\\ {{\it 1 Jan Smuts Ave, Johannesburg 2000, South Africa}}
  \spa{0.5} \\
$^b${{\it Department of Physics, Nagoya University}}
\\ {{\it Furo-cho, Chikusa-ku, Nagoya 464-8602, Japan}}
\spa{0.5}  \\
\&
\spa{0.5}  \\
$^c${{\it Center for Gravitational Physics}}
\\ {{\it  Yukawa Institute for Theoretical Physics, Kyoto University}}
\\ {{\it Kitashirakawa-Oiwakecho, Sakyo-ku, Kyoto 606-8502, Japan}}
\spa{0.5}  

\end{center}

\begin{abstract}
We study the random geometry approach to the $T\bar{T}$ deformation of
2$d$ conformal field theory developed by Cardy and discuss its
realization in a gravity dual.  In this representation, the gravity dual
of the $T\bar{T}$ deformation becomes a straightforward translation of
the field theory language.  Namely, the dual geometry is an ensemble of
AdS$_3$ spaces or BTZ black holes, without a finite cutoff, but instead
with randomly fluctuating boundary diffeomorphisms.  This reflects an increase
in degrees of freedom in the renormalization group flow to the UV by the
irrelevant $T\bar{T}$ operator. We streamline the method of computation  
and calculate the energy spectrum and the thermal free energy in a
manner that can be directly translated into the gravity dual
language. We further generalize this approach to correlation functions
and reproduce the all-order result with universal logarithmic
corrections computed by Cardy in a different method. In contrast to earlier proposals, 
this version of the gravity dual of the $T\bar{T}$ deformation works not only for the
energy spectrum and the thermal free energy but also for correlation
functions.
\end{abstract}

\renewcommand{\thefootnote}{\arabic{footnote}}
\setcounter{footnote}{0}

\newpage

\tableofcontents


\section{Introduction}
\label{Sec:Introduction}

Irrelevant deformations in quantum field theory (QFT) are, in general, hard to be tamed due to power-counting non-renormalizability. Even though they are sensible within the effective field theory framework and play important roles in the phenomenological model building, only few of them may be consistent with UV-completions of low energy effective theories.

\medskip
It thus came as a surprise that the so-called ``$T\bar{T}$'' deformation \cite{Zamolodchikov:2004ce}, that universally exists in any 2$d$ QFTs except for exotic theories without an energy-momentum tensor, turned out to be UV-complete and, moreover, preserve integrability \cite{Smirnov:2016lqw}. 
In contrast to asymptotic safety \cite{Weinberg:1978, Weinberg:1980gg}, however, the $T\bar{T}$-deformed theories do not flow to UV fixed points and exhibit signs of non-locality  \cite{Dubovsky:2012wk, Dubovsky:2017cnj}.
Meanwhile, owing to integrability, the otherwise-challenging energy spectrum problem, for instance, can be solved exactly \cite{Smirnov:2016lqw, Cavaglia:2016oda}.
Thus despite the deformation by an oft-uncontrollable irrelevant operator, the $T\bar{T}$-deformed theories are remarkably tractable and marginally well-behaved.\footnote{A recent work \cite{Haruna:2020wjw} studied a large $N$ free $O(N)$ vector model and reported that a very higher-energy singlet mode, which can be identified with a conformal mode, has a negative norm indicating that the $T\bar{T}$-deformed theories are non-unitary.} 
To study further and better understand the $T\bar{T}$-deformed theories, it would be desirable to demystify these somewhat surprising features of the $T\bar{T}$ deformation and make the underlying simplicity manifest. Indeed, such a formulation was developed by Cardy \cite{Cardy:2018sdv}.
It is a particular representation of the $T\bar{T}$ deformation via a Hubbard-Stratonovich transformation in which the deformation can be regarded, essentially, as random coordinate transformations.
A somewhat related approach was proposed in \cite{Dubovsky:2017cnj, Dubovsky:2018bmo} where it was demonstrated that the $T\bar{T}$ deformation can be thought of as coupling a QFT to the JT gravity \cite{Jackiw:1984je, Teitelboim:1983ux} in the flat space limit.
Here we build our work on the former ``random geometry approach'' developed by Cardy. 

\medskip
Our objectives are twofold: One is to study further Cardy's random geometry approach  to the  $T\bar{T}$ deformation. The other is to discuss its gravity dual realization. In the course of discussions, we streamline the method of computation so that the gravity dual description becomes merely a straightforward translation of languages from field theory to gravity. We then generalize Cardy's original formulation to the case with local operator singularities so that our method can be applied to the computation of correlation functions in the $T\bar{T}$-deformed theories. In our new streamlined method we rederive the energy spectrum \cite{Smirnov:2016lqw, Cavaglia:2016oda}, the thermal free energy \cite{Cardy:2018sdv} and correlation functions \cite{Cardy:2019qao, Kraus:2018xrn} computed in different methods.

\medskip
In this representation of the $T\bar{T}$ deformation, as a straightforward deductive logic, the gravity dual is  an ensemble of AdS$_3$ spaces or BTZ black holes, without a finite cutoff, but instead with randomly fluctuating boundary diffeomorphisms. This bodes well with the fact that the $T\bar{T}$ operator is geometric and irrelevant: Being irrelevant, the $T\bar{T}$ operator drives a renormalization group flow to the UV and new degrees of freedom are integrated in over the scale of the $T\bar{T}$ coupling $\mu$. On the one hand, this brings in a novel concept to the AdS/CFT framework. 
On the other hand, this may be too straightforward and anticlimactic. However, in contrast to the earlier proposals, i.e.~the cutoff AdS by McGough-Mezei-Verlinde \cite{McGough:2016lol} and the mixed nonlinear boundary condition by Guica-Monten \cite{Guica:2019nzm}, it is worth emphasizing that our version of the gravity dual works not only for the energy spectrum and the thermal free energy but also for correlation functions.

\medskip
Finally, although it is rather a digression, we would like to add a few words on our initial motivation. One well-known example of irrelevant deformations in AdS$_{d+1}$/CFT$_d$ is the deformation by a dimension $2d$ operator that corresponds to moving away from the near-horizon limit to an asymptotically flat spacetime \cite{Gubser:1998kv, Intriligator:1999ai}. 
It was further speculated in \cite{Intriligator:1999ai} that the dual supersymmetric Yang-Mills theory is deformed to the Dirac-Born-Infeld (DBI) theory.
Indeed, quadratic operators composed of the stress tensor generically have conformal dimension $2d$ and are purely geometric in the sense of supergravity multiplet. Furthermore, in the $d=2$ case, the deformed field theory is of the Nambu-Goto type \cite{Dubovsky:2012wk, Caselle:2013dra, Cavaglia:2016oda, Callebaut:2019omt} analogous to the DBI theory. 
It is then natural to ask if the $T\bar{T}$ deformation can be interpreted as a deformation to an asymptotically flat spacetime. 
In fact, this is the idea behind the work \cite{Giveon:2017nie} which interprets a deformation of an AdS$_3$ space to an asymptotically flat linear dilaton spacetime as a type of $T\bar{T}$ deformations. However, the ``$T\bar{T}$'' operator in this case is typically single-trace as opposed to double-trace and thus differs from the widely-discussed ``$T\bar{T}$'' operator considered in \cite{Zamolodchikov:2004ce}.\footnote{We thank Ofer Aharony and Sam van Leuven for discussions on this point. We also note that correlation functions in the $SL(2,\mathbb{R})$ WZW model deformed by a boundary single-trace ``$T\bar{T}$'' operator were computed in \cite{Giribet:2017imm, Babaro:2018cmq, Giribet:2020kde}.}  
With hindsight, pure Einstein gravity with a negative cosmological constant does not admit any asymptotically flat spacetimes but only allows AdS$_3$ space or its quotients, BTZ black holes. Thus this excludes outright a possibility of connecting the standard  $T\bar{T}$ deformation to moving away from the near-horizon limit.

\medskip
The organization of the paper is as follows: We start with a brief review of Cardy's random geometry approach to the $T\bar{T}$ deformation in Section \ref{Sec:RandomGeometry}. We then outline a schematic formulation of  the gravity dual description based on the random geometry approach in Section \ref{Sec:GravityDual}.
To illustrate how exactly this framework works, in Section \ref{Sec:EnergySpectrum}, we consider concrete examples and give detailed computations of the energy spectrum and the thermal free energy. The field theory analysis for each example is followed by a straightforward translation to the gravity dual description. We also discuss a relation between two examples from the perspective of ensembles, i.e.~microcanonical vs.\ canonical. 
We then move on to the discussion of correlation functions on $\mathbb{R}^2$ in Section \ref{Sec:Correlators}. In order to compute correlation functions in the random geometry approach, we first generalize Cardy's argument in Section  \ref{Sec:RandomGeometry} to the case with local operator singularities. In our new method, not only do we reproduce the results found in the earlier works \cite{Cardy:2019qao, Kraus:2018xrn}, but it also becomes very straightforward how the field theory computation can be replicated in the gravity dual.
In Section \ref{Sec:Discussions}, we briefly summarize the main results of this work and give short discussions on several future directions of our interest. 
Finally, some details of the gravity computations are relegated to Appendices \ref{Appendix:ADMmass} and \ref{Appendix:Onshellaction}.


\section{$T\bar{T}$ deformation as random geometry}
\label{Sec:RandomGeometry}

An illuminating view of the $T\bar{T}$ deformation was suggested and developed by Cardy \cite{Cardy:2018sdv}, refining his own earlier idea \cite{Cardy:2015xaa}.\footnote{A very similar idea was proposed in \cite{McGough:2016lol} using the results in \cite{Freidel:2008sh} that is based on the work \cite{Verlinde:1989ua}. It is a finite version of the infinitesimal Hubbard-Stratonovich transformation discussed in \cite{Cardy:2015xaa}.
However, it appears that these two ideas are not the same and differ in important ways, as will be elaborated below.}
This is the key perspective as well as the main technical tool we adopt in this work. We thus provide a brief review of the main points in Cardy's random geometry approach.

\medskip
The ``$T\bar{T}$'' operator ${\cal O}_{T\bar{T}}$ is defined by 
\be
 \label{TTbaroperator}
 {\cal O}_{T\bar{T}}\equiv T\bar{T}-\Theta^2
 = -{1\over 8}\epsilon_{ik}\epsilon_{jl}T^{ij}T^{kl}
 = -{1\over 4}\det T_{ij}\ ,
\ee
where $T={1\over 4}(T_{11}-T_{22}-2iT_{12})$, $\Theta={1\over 4}(T_{11}+T_{22})$, and $i,j,...=1,2$.
We define the stress-energy tensor via the variation of the Euclidean action:
\be
\delta S=-\int d^2x\,\sqrt{g}\,T^{ij}\delta g_{ij}.
\label{Tij_def}
\ee

\medskip
A $T\bar{T}$-deformed theory ${\cal T}[\mu]$ is characterized by a finite coupling $\mu$ of length dimension two. Then the $T\bar{T}$ deformation is best defined as an infinitesimal change of the action from ${\cal T}[\mu]$ to ${\cal T}[\mu+\delta\mu]$:
\be
\label{DefineDeformedTheory}
S[\mu+\delta\mu]=S[\mu]+\delta\mu\int d^2x\,{\cal O}_{T\bar{T}}\equiv S[\mu]+\delta S\ .
\ee
An important  note is that the stress tensor $T_{ij}$, composing the $T\bar{T}$ operator, is that of the deformed theory ${\cal T}[\mu]$ rather than that of the undeformed theory ${\cal T}[0]$. The deformed theory ${\cal T}[\mu]$ of a finite coupling $\mu$ can be constructed from the undeformed theory ${\cal T}[0]$ by iteration of infinitesimal deformations \eqref{DefineDeformedTheory}.

\medskip
Now, the idea is to split $T\bar{T}$ by a Hubbard-Stratonovich transformation
\be
\label{actionchange}
\exp\left(-\delta S\right)\propto \int [dh]\exp\biggl[-{1\over 8\delta\mu}\int d^2x\,\epsilon^{ik}\epsilon^{jl}h_{ij}h_{kl}+\int d^2x\, h_{ij}T^{ij}\biggr]\ .
\ee
There are two important remarks to be made: (1) For an infinitesimal $\delta\mu$, the $h_{ij}$-integrals are dominated by the saddle point;
\begin{align}\label{Sec2:saddlepoint}
{1\over 4\delta\mu}h_{22}^{\ast}=T_{11}\ , \qquad {1\over 4\delta\mu}h_{11}^{\ast}=T_{22}\ ,\qquad
{1\over 4\delta\mu}h_{12}^{\ast}=-T_{12}\ ,\qquad {1\over 4\delta\mu}h_{21}^{\ast}=-T_{21}\ .
\end{align}
(2) Since the stress tensor $T^{ij}$ is by definition a response to a small change of the metric, this form of the action implies that the $T\bar{T}$ deformation can be thought of as random changes of the background metric $g_{ij}\to g_{ij}+h_{ij}$.  

\medskip
This, however, is not the end of the story and this is where the idea in  \cite{Cardy:2018sdv} differs most from the earlier works \cite{Cardy:2015xaa} and \cite{McGough:2016lol, Freidel:2008sh, Verlinde:1989ua}. An obvious but important fact is that the stress tensor is conserved in the absence of local operator singularities:\footnote{When local operators are inserted, the conservation law has the $\delta$-function sources. As we discuss in Section \ref{Sec:Correlators}, it is crucial to take them into account for the computation of correlators. }\label{singularities}
\begin{align}
\del_iT^{ij}=0
\end{align}
which imposes constraints on the saddle point
\begin{align}\label{Hconservation}
\del_1h_{22}^{\ast}=\del_2h_{21}^{\ast}\ ,\qquad\qquad \del_2h_{11}^{\ast}=\del_1h_{12}^{\ast}\ .
\end{align}
These can be solved by
\be\label{curlfreediffeo}
h_{ij}=\del_i\alpha_j+\del_j\alpha_i\qquad\quad\mbox{with}\qquad\quad \epsilon^{ij}\del_i\alpha_j=0\ .
\ee
This means that the effect of the $T\bar{T}$ deformation merely amounts to (curl-free) coordinate transformations 
\be\label{coordinatetransform}
x_i\mapsto x_i +\alpha_i
\ee
rather than general changes of the metric.  $\alpha_i$ can be non-single-valued, although $h_{ij}$ must be single-valued.
An important implication is that the saddle point action becomes a total derivative
\begin{align}\label{saddleactiontotdel}
-\delta S^{\ast}={1\over 8\delta\mu}\int_{\cal M} d^2x\,\epsilon^{ik}\epsilon^{jl}h^{\ast}_{ij}h^{\ast}_{kl}
={1\over 2\delta\mu}\int_{\del{\cal M}}ds^{k}\epsilon^{jl}\alpha_j\del_k\alpha_l 
\end{align}
where we used the curl-free condition and $ds^k$ is a vector tangent to the boundary $\del{\cal M}$ or nontrivial cycles as in the case of a 2-torus $T^2$ or a cylinder $\mathbb{R}\times S^1$.   

\medskip
Hence, in the absence of local operator singularities, there can be
effects of the $T\bar{T}$ deformation only if the 2$d$ manifold has
boundaries or nontrivial cycles. However, as commented in footnote
\ref{singularities}, when local operators are inserted, there are
effects of the $T\bar{T}$ deformation from singularities even if the
manifold is a topologically trivial infinite plane $\mathbb{R}^2$.  (Or,
alternatively, we can say that, upon removing points of insertions,
$\mathbb{R}^2$ has gained nontrivial topology.)


\section{Gravity dual}
\label{Sec:GravityDual}

In the random geometry approach, the gravity dual becomes a straightforward translation of the field theory language. 
First, in the field theory language, since the $T\bar{T}$ deformation is equivalent to random changes of the background metric $g_{ij}\to g_{ij}+h_{ij}$ constrained to \eqref{curlfreediffeo}, a physical observable ${\cal O}_{{\cal T}[\delta\mu]}\left[\{x\}\right]$ of an infinitesimally deformed theory ${\cal T}[\delta\mu]$ is obtained from a physical observable ${\cal O}_{{\cal T}[0]}\left[\{x\}\right]$ of a undeformed theory ${\cal T}[0]$ via
\begin{align}\label{FTdeltamu}
{\cal O}_{{\cal T}[\delta\mu]}\left[\{x\}\right]={\cal N}^{-1}
\int [dh]\exp\biggl[-{1\over 8\delta\mu}\int d^2x\,\epsilon^{ik}\epsilon^{jl}h_{ij}h_{kl}\biggr]{\cal O}_{{\cal T}[0]}\left[\{x+\alpha\}\right]
\end{align}
where the normalization factor ${\cal N}$ is the Gaussian integrals of $h_{ij}$, and $\{x\}$ collectively denotes any coordinate dependence including the coordinate lengths.  

\medskip
Translating this formula into the gravity dual language, 
we start with the genuine AdS/CFT without any deformations \cite{Maldacena:1997re} corresponding to the undeformed theory ${\cal T}[0]$. Then we add new degrees of freedom, i.e.~boundary metric deformations $h_{ij}$, restricted to diffeomorphisms \eqref{curlfreediffeo}. They randomly fluctuate over a scale $\delta\mu$.   
This bodes well with the fact that the $T\bar{T}$ operator is geometric and irrelevant. Being irrelevant means that  the $T\bar{T}$ operator drives a renormalization group flow to the UV over the scale $\delta\mu$ and thus new degrees of freedom are integrated in over this scale.

\begin{figure}[h!]
\centering
\centering \includegraphics[height=2.6in]{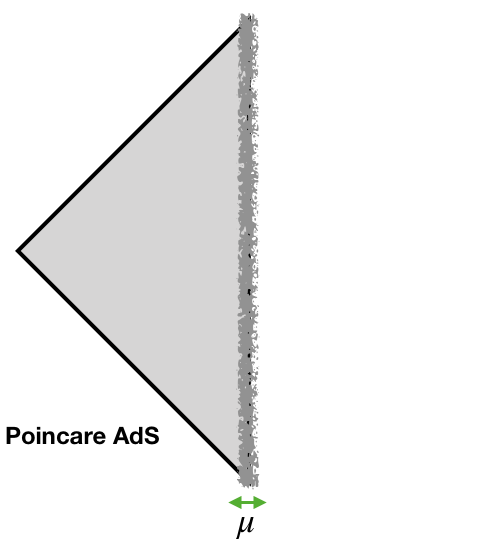}
\hspace{1cm}\vspace{0cm}
\centering \includegraphics[height=2.2in]{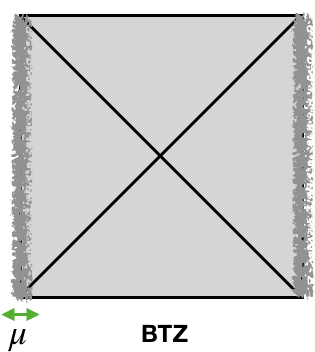}
\caption{The gravity dual of $T\bar{T}$ deformation: AdS$_3$ and BTZ black hole with random boundary geometry with the metric $h_{ij}$ that fluctuates over the scale $\mu$. The gravity dual is an ensemble of AdS$_3$ spaces or BTZ black holes.}
\label{fig:PoincareAdS}
\end{figure}  

\medskip
In the standard GKP-W dictionary \cite{Gubser:1998bc, Witten:1998qj}, 
the factor $\exp\left(\int d^2x\, h_{ij}T^{ij}\right)$ in the deformation of the field theory action \eqref{actionchange} corresponds to turning on a non-normalizable mode of the metric. As has already been stressed, the non-normalizable mode in this case is constrained to the one generated by  boundary diffeomorphisms, and thus we only need to apply (curl-free) boundary coordinate transformations \eqref{coordinatetransform}. Note that this is merely a holographic rephrasing of the aforementioned field theory statement. Hence, given the undeformed AdS/CFT correspondence 
\be
{\cal O}_{\rm AdS}\left[g^{\rm bdy}_{ij}(x), \phi^{\rm bdy}(x)\right]={\cal O}_{{\cal T}[0]}\left[\{x\}\right]\ ,
\ee
albeit rather anticlimactic, any quantity in the gravity dual ${\rm AdS}[\delta\mu]$ of the ${\cal T}[\delta\mu]$ theory can be computed as
\begin{align}\label{AdSdeltamu}
{\cal O}_{{\rm AdS}[\delta\mu]}\left[g^{\rm bdy}_{ij}(x), \phi^{\rm bdy}(x)\right]=&\,{\cal N}^{-1}
\int [dh]\exp\biggl[-{1\over 8\delta\mu}\int d^2x\,\epsilon^{ik}\epsilon^{jl}h_{ij}h_{kl}\biggr]\nn\\
&\hspace{3cm}\times {\cal O}_{\rm AdS}\left[g^{\rm bdy}_{ij}(x+\alpha), \phi^{\rm bdy}(x+\alpha)\right]\ ,
\end{align}
where $g^{\rm bdy}_{ij}(x)$ and $\phi^{\rm bdy}(x)$ are the boundary values of the (non-radial) metric and any other fields, respectively.
The gravity dual ${\rm AdS}[\mu]$ for a finite $\mu$ can be constructed by iteration of infinitesimal deformations \eqref{AdSdeltamu}.

\medskip
In the following sections, we will demonstrate how this framework precisely works, filling in schematic and oversimplified parts of the above formulas \eqref{FTdeltamu} and \eqref{AdSdeltamu}.


\section{Energy spectrum and thermal free energy}
\label{Sec:EnergySpectrum}

As an application of the random geometry approach to the $T\bar{T}$ deformation, we first compute the energy spectrum \cite{Smirnov:2016lqw, Cavaglia:2016oda} and the thermal free energy \cite{Cardy:2018sdv}. Although the results are well-known, we streamline the method of computation in a manner that can be straightforwardly translated into the gravity dual language. Thus our exposition of the calculation offers a more direct and simpler way than that in \cite{Cardy:2018sdv}.

\begin{figure}[h!]
\centering
\centering 
\begin{tikzpicture}  
\draw[thick] (-2,0) ellipse (.3 and 1);
\draw[thick] (+2,0) ellipse (.3 and 1);
\draw[thick] (-2, 1) -- (2, 1);
\draw[thick] (-2,-1) -- (2,-1);
\draw[latex-latex] (-2,1.2) -- (2,1.2) node [midway,above] () {$L$};

\draw[densely dotted] (2,0) -- +(-75:-0.75);
\draw[latex-latex] (2,0) -- +(-75:0.75) node [midway,right,xshift=2] () {$R$};

\draw[-latex] (-1,-1.5) -- (1,-1.5) node [right] () {$x_1$};

 \draw[-latex] (3,-.2) arc (-160:170:0.15 and 0.5) node [right,midway,xshift=0] {$x_2$};
\end{tikzpicture}
\caption{A finite cylinder of length $L$ and radius $R$.}
\label{fig:cylinder}
\end{figure}
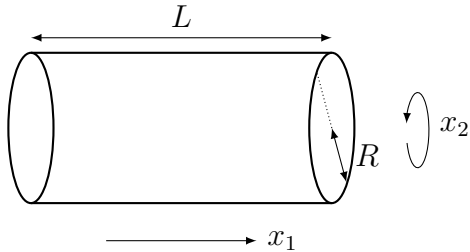

\medskip
To analyze the energy spectrum and the thermal free energy, we consider a finite cylinder $I\times S^1$ where an open interval $I=\{x_1\,|\,  0\le x_1\le L\}$  and $S^1$ is a circle of radius $R$, i.e.~$S^1=\{x_2\,|\,  0\le x_2\le 2\pi R\}$ as illustrated in Figure \ref{fig:cylinder}. 

Even though the random metric $h_{ij}$ is constrained to diffeomorphisms \eqref{curlfreediffeo} as discussed in Section \ref{Sec:RandomGeometry}, for practical purposes, it is more convenient  in this particular case to work with the metric $h_{ij}$. As argued in \cite{Cardy:2018sdv}, the diffeomorphisms \eqref{curlfreediffeo} together with the property of the saddle point action \eqref{saddleactiontotdel} imply that the $h_{ij}$-integrals are localized on \emph{constant} metrics $h_{11}$ and $h_{22}$ with $h_{12}=h_{21}=0$.
The argument goes as follows: We think of a finite cylinder as a rectangle with width $L$ and height $2\pi R$ with the top and bottom edges being identified.
The saddle point action is a total derivative \eqref{saddleactiontotdel} and given in this case by
\begin{align}
\hspace{-.3cm}
-\delta S^{\ast}={1\over 2\delta\mu}\Biggl[\left.\int_0^{2\pi R}dx_2\left(\alpha_1\del_2\alpha_2-\alpha_2\del_2\alpha_1\right) \right|_{x_1=0}^{x_1=L}
-\left.\int_0^{L}dx_1\left(\alpha_1\del_1\alpha_2-\alpha_2\del_1\alpha_1\right) \right|_{x_2=0}^{x_2=2\pi R}\Biggr].
\end{align}
A nontrivial winding or discontinuity of $\alpha_i$ is allowed along a nontrivial cycle $S^1$ in the vertical $x_2$-direction, but the metric has to be single-valued, i.e.~taking the same value at the top and bottom edges. This implies that $\alpha_1=\half h_{21}x_2+\tilde{\alpha}_1(x_1,x_2)$ and $\alpha_2=\half h_{22}x_2+\tilde{\alpha}_2(x_1,x_2)$ with $\tilde{\alpha}_i(x_1,x_2+2\pi R)=\tilde{\alpha}_i(x_1,x_2)$. The curl-free condition imposes $\half h_{21}+\del_2\tilde{\alpha}_1=\del_1\tilde{\alpha}_2$.
It is then straightforward to show that\footnote{This can be most easily understood by Fourier-expanding $\tilde{\alpha}_i(x_1,x_2)=\sum_{n=0}^{\infty}(\cos\left(n x_2/R\right)\tilde{\alpha}^c_{n,i}(x_1)+\sin\left(n x_2/R\right)\tilde{\alpha}^s_{n,i}(x_1))$.}
\begin{align}
\hspace{-.3cm}
-\delta S^{\ast}&=-{1\over\delta\mu}\left.\int_0^{2\pi R}dx_2\tilde{\alpha}_2\del_1\tilde{\alpha}_2 \right|_{x_1=0}^{x_1=L}
+{\pi R\over \delta\mu}\left(\tilde{\alpha}_1(L, 2\pi R)-\tilde{\alpha}_1(0, 2\pi R)\right) h_{22}\nn\\
&=\left.2\int_0^{2\pi R}dx_2\tilde{\alpha}_2 T_{12}\right|_{x_1=0}^{x_1=L}+{2\pi RL\over 4\delta\mu}h_{11}h_{22}\ ,
\end{align}
where we used $\del_1\tilde{\alpha}_2=-2\delta\mu\, T_{12}$ and the curl-free condition and that the change of the interval length $\delta L\equiv \tilde{\alpha}_1(L, 2\pi R)-\tilde{\alpha}_1(0, 2\pi R)$ can be regarded as a scaling reparametrization of  $x_1$ by a constant metric $\delta L=(\sqrt{1+h_{11}}-1)L=h_{11}L/2+{\cal O}(\delta\mu^2)$.
Now, we impose the boundary condition that there is no momentum flow out of the edges of the cylinder, $T_{12}(L, x_2)=T_{12}(0,x_2)=0$.
Then the saddle point action depends only on constant $h_{11}$ and $h_{22}$:
\begin{align}
\delta S^{\ast}=-{2\pi RL\over 4\delta\mu}h_{11}h_{22}\ .
\end{align}
Thus the physical observables in the ${\cal T}[\delta\mu]$ theory can be computed as
\begin{align}\label{deformcylinder}
{\cal O}_{{\cal T}[\delta\mu]}\left[g_{ii}=1\right]={\cal N}^{-1}
\int [dh]\exp\left[-{2\pi R L\over 4\delta\mu}h_{11}h_{22}\right]{\cal O}_{{\cal T}[0]}\left[g_{ii}=1+h_{ii}\right]\ .
\end{align}
Note that the Gaussian action is the one without the linear $hT$ term as implied in the above formula \eqref{FTdeltamu} and \eqref{AdSdeltamu} and thus minus the saddle point action.


\subsection{Energy spectrum}
\label{Sec:ES}

The energy spectrum of the $T\bar{T}$-deformed theory has been first computed in \cite{Smirnov:2016lqw, Cavaglia:2016oda}. Although it was reproduced by the random geometry approach in  \cite{Cardy:2018sdv}, we provide a more direct and simpler way of computation than that presented in  \cite{Cardy:2018sdv}.
We work in  Lorentzian signature and thus Wick-rotate back the coordinate $x_1$ to real time $t$, i.e.~$L=it$.


\subsubsection{Field theory}
\label{subsecES:FT}

The object of interest is
the contribution to the (Lorentzian) partition function from an energy eigenstate:
\begin{align}\label{eigenenergyPF}
z_n(t, R)\equiv\langle n|e^{-i\hat{H}(R)t}|n\rangle=e^{-iE_n(R)t}
\end{align}
which depends on the metric $g_{ij}$ through the coordinate lengths $t$
and $R$.  Applying the formula \eqref{deformcylinder}, the energy
spectrum of the ${\cal T}[\delta\mu]$ theory can be computed as
\begin{align}\label{ES1storder}
z_n(t, R; \delta\mu)&= {\cal N}^{-1}\int dh\, e^{-it{\pi R\over 2\delta\mu}h_{11}h_{22}}
\,z_n\!\left(\sqrt{1+h_{11}}t, \sqrt{1+h_{22}}R\right)
\equiv {\cal N}^{-1}\int dh \,e^{-i{\cal E}[h]t}\ ,
\end{align}
where the deformed energy spectrum $E_n(R; \delta\mu)$  is naturally defined through
\be
z_n(t, R; \delta\mu)=f_n(R;\delta\mu)\,e^{-i E_n(R; \delta\mu) t}\ ,
\ee
where $f_n(R;\delta\mu)$ is a time-independent factor. 
The undeformed energy of a primary state with conformal weight $(\Delta_n, \bar{\Delta}_n)$ is given by
\be\label{CFTenergy}
E_n(R)={\Delta_n+ \bar{\Delta}_n-{c\over 12}\over R}\equiv {C_n\over R}\ .
\ee
The $h$-integrals in \eqref{ES1storder} can be well approximated by the saddle point for an infinitesimal~$\delta\mu$. The explicit form of the exponent is
\begin{align}
{\cal E}[h]&={\pi R\over 2\delta\mu}h_{11}h_{22}+\sqrt{1+h_{11}}E_n\left(\sqrt{1+h_{22}}R\right)\nn\\
&={C_n\over R}+{\pi R\over 2\delta\mu}h_{11}h_{22}+\half(h_{11}-h_{22}){C_n\over R}+{\cal O}(\delta\mu^2)
\label{ES1storderv2}
\end{align}
whose saddle point is at
\be
h^{\ast}_{11}=-h^{\ast}_{22}={\delta\mu\, C_n\over \pi R^2}\ .
\ee
Thus the energy at the saddle point is found to be
\be
E_n(R; \delta\mu)={\cal E}[h^{\ast}]={C_n\over R}+{\delta\mu\, C_n^2\over 2\pi R^3}\ .
\ee
This reproduces the first-order correction in the all-order energy spectrum \cite{Smirnov:2016lqw, Cavaglia:2016oda, Cardy:2018sdv}
\be\label{SZformula}
E_n(R; \mu)={\pi R\over \mu}\left[1-\sqrt{1-{2\mu C_n\over \pi R^2}}\right]\ .
\ee
To be complete, the fluctuations about the saddle point must be taken
into account. This involves keeping ${\cal O}(h^2)$ terms in the
expansion in \eqref{ES1storderv2}.  However, it is easy to see that the
Gaussian integration over the fluctuations only yields a
time-independent factor $f_n(R;\delta\mu)$ and thus does not contribute
to the energy spectrum.


\paragraph{$\bullet$ All orders I -- PDE (Burgers' equation)}

Now, when the above procedure \eqref{ES1storderv2} is applied to an infinitesimal deformation from the ${\cal T}[\mu]$ theory to the ${\cal T}[\mu+\delta\mu]$ theory, it becomes
\begin{align}\label{ESfiniteorder}
E_n(R;\mu+\delta\mu)={\pi R\over 2\delta\mu}h^{\ast}_{11}h^{\ast}_{22}+\Bigl(1+\half h^{\ast}_{11}\Bigr)E_n(R;\mu)
+\half h^{\ast}_{22}R\,\del_R E_n(R;\mu)+{\cal O}(\delta\mu^2)\ .
\end{align}
where the saddle point is at
\be\label{Burgerssaddle}
h^{\ast}_{11}=-{\delta\mu \over \pi }\del_RE_n(R;\mu)\ ,\qquad\qquad
h^{\ast}_{22}=-{\delta\mu \over \pi R}E_n(R;\mu)\ .
\ee
Thus the equation \eqref{ESfiniteorder} for an infinitesimal deformation yields
\begin{align}\label{Burgers}
\del_{\mu}E_n(R;\mu)=-{1\over 2\pi }E_n(R;\mu)\,\del_RE_n(R;\mu)\ .
\end{align}
This is indeed the inviscid Burgers equation with a driving force, which
is obeyed by the energy spectrum of the $T\bar{T}$-deformed theory
\cite{Smirnov:2016lqw, Cavaglia:2016oda}.  This means that we have
reproduced the all-order energy spectrum \eqref{SZformula}.  As a
further remark, we note that
\begin{align}
T_i^i=-{1\over 4\pi R}\del_R\left(RE_n(R;\mu)\right)
=-{\mu\over (2\pi)^2R }E_n(R;\mu)\,\del_R E_n(R;\mu)
=-4\mu\det T_{ij}\label{floweq_trT_detT}
\end{align}
where $T_i^i=T_{11}+T_{22}$.  In the first and third equalities we used
\eqref{Sec2:saddlepoint} and \eqref{Burgerssaddle}, while in the second
equality we used
$\del_R(RE_n(R;\mu))=-2(\mu/R)\,\del_{\mu}(RE_n(R;\mu))$
which follows on dimensional grounds.
This equation \eqref{floweq_trT_detT}
is the flow equation that appeared in \cite{Smirnov:2016lqw,
Cavaglia:2016oda}.


\paragraph{$\bullet$ All orders II -- order-by-order iteration}

An alternative way to compute higher order corrections is to iterate
infinitesimal deformations order by order by power expansion in $\mu$. A key idea is that
the energy spectrum at the $i$-th order can be expressed as
\be
E_n^{(i)}(R;\mu)={C_n\over R^{(i)}}+{\cal O}(\mu^{i+1})\ ,
\ee
where the radius $R^{(i)}$ at the $i$-th order is given by
\be
R^{(i)}=R\left(1+{1\over 2}h_{22}^{\ast(i)}\right)\simeq R\sqrt{1+h_{22}^{\ast(i)}}\ .
\ee
The superscript ${(i)}$ means an expression that contains powers of $\mu$ up to $\mu^i$.
Then the higher order generalization of  \eqref{ES1storderv2} reads
\begin{align}
E^{(i)}_n(R; \mu)
&=E_n(R)+{\pi R\over 2\mu}h^{\ast(i)}_{11}h^{\ast(i)}_{22}+\half\left(h^{\ast(i)}_{11}-h^{\ast(i)}_{22}\right)E^{(i-1)}_n\left(R;\mu\right)
+{\cal O}(\mu^{i+1})
\end{align}
whose saddle point is at
\be\label{iteration:saddle}
h^{\ast(i)}_{11}=-h^{\ast(i)}_{22}={\mu\over \pi R}E^{(i-1)}_n\left(R;\mu\right)\ .
\ee
Thus we obtain the recursion relation
\begin{align}\label{recursion}
E^{(i)}_n(R; \mu)=E_n(R)+\left[{\mu\over 2\pi R}\left(E^{(i-1)}_n\left(R;\mu\right)\right)^2\right]_{i}\ ,
\end{align}
where the symbol $[X(\mu)]_{i}$ denotes that $X(\mu)$ as a series in $\mu$ is truncated at ${\cal O}(\mu^i)$. The continuum version of this recursion relation is given by
\be
E_n(R; \mu)=E_n(R)+{\mu\over 2\pi R}\left(E_n(R; \mu)\right)^2\ .
\ee
This is indeed solved by the all-order energy spectrum \eqref{SZformula}.

\medskip
As a remark, the saddle point \eqref{iteration:saddle} implies that the deformed energy spectrum can be thought of as a deformation of the radius
\be
R^{(i)}=R\left(1-{\mu\over 2\pi R}E^{(i-1)}_n\left(R;\mu\right)\right)
\ee
in the undeformed CFT energy, as observed in \cite{Smirnov:2016lqw}. Furthermore, due to this rescaling of the radius in the $x_2$-direction, the $22$-component of the stress tensor of the deformed theory is transformed to
\be
T_{22}^{(i)}\quad\mapsto\quad \tilde{T}_{22}^{(i)}={h^{\ast(i)}_{11}\over 4\mu}\left(1-h_{22}^{\ast(i)}\right)
=-{h^{\ast(i)}_{22}\over 4\mu}+\mu\left({1\over 2\pi R}E^{(i-1)}_n\left(R;\mu\right)\right)^2\ ,
\ee
where a tilde denotes quantities in the deformed theory.
This then yields
\be
\tilde{T}_i^i=T_{11}+\tilde{T}_{22}=-4\mu\det T_{ij}\ ,
\ee
where we dropped the superscript $(i)$ to avoid clutter of notation. 
This is again the flow equation in  \cite{Smirnov:2016lqw, Cavaglia:2016oda}.


\subsubsection{Gravity dual}
\label{subsecES:GD}

As repeatedly stated, in this approach, the gravity dual description of
the $T\bar{T}$ deformation becomes a straightforward translation of the
field theory language.  
The contribution to the partition function from an energy eigenstate, \eqref{eigenenergyPF},
is translated into
\be
z_n(g^{\rm bdy}_{ij})=e^{-i H_{\rm BTZ}(g^{\rm bdy}_{ij}) t}
\ee
where $H_{\rm BTZ}$ is the Hamiltonian of the pure Einstein gravity with negative cosmological constant evaluated on (non-rotating) BTZ black holes \cite{Banados:1992wn}.
As reviewed in \cite{Brown:1994gs}, Brown and York have shown that the Hamiltonian has a nonvanishing contribution from the boundary~\cite{Brown:1992br} which in this case is the ADM mass of the BTZ black holes,
\be
H_{\rm BTZ}(g^{\rm bdy}_{ij})=M_{\rm ADM}\ .
\ee
As shown in detail in Appendix \ref{Appendix:ADMmass}, for the BTZ black holes 
\begin{align}\label{nonrotateBTZ}
ds_{\rm BTZ}^2=-\left(\frac{\rho ^2}{4}-\frac{M}{2}+\frac{M^2}{4 \rho ^2}\right)dt^2+\left(\frac{\rho ^2}{4}+\frac{M}{2}+\frac{M^2}{4 \rho ^2}\right)dy^2+{d\rho^2\over\rho^2}
\end{align}
with $y\sim y+2\pi R$, the ADM mass is given by $M_{\rm ADM}=RM$.
In the AdS/CFT dictionary, this is identified with the CFT energy \eqref{CFTenergy}, $M_{\rm ADM}=C_n/R$, and 
thus the ``mass'' $M$ in the metric \eqref{nonrotateBTZ} is more properly parametrized by
\be
M={C_n\over R^2}\ .
\ee
This implies that under a diffeomorphism induced by the $T\bar{T}$ deformation, the BTZ black holes are deformed to
\be\label{BTZdeformation}
g_{tt}\mapsto (1+h_{11})\,g_{tt}\ ,\qquad g_{yy}\mapsto (1+h_{22})\,g_{yy}\ ,\qquad M\mapsto M/(1+h_{22})\ .
\ee 
Then the ADM mass is transformed to
\be
M_{\rm ADM}\mapsto \sqrt{1+h_{11}\over 1+h_{22}}M_{\rm ADM}\ .
\ee
Thus rephrasing Burgers' equation \eqref{Burgers} and the recursion equation \eqref{recursion} in terms of the deformed ADM mass $M_{\rm ADM}(R;\mu)$, we simply obtain that
\begin{align}
\del_{\mu}M_{\rm ADM}(R;\mu)&=-{1\over 2\pi }M_{\rm ADM}(R;\mu)\,\del_RM_{\rm ADM}(R;\mu)\ ,\label{GDBurgers}\\
M_{\rm ADM}^{(i)}(R; \mu)&=M_{\rm ADM}+\biggl[{\mu\over 2\pi R}\left(M_{\rm ADS}^{(i-1)}\left(R;\mu\right)\right)^2\biggr]_{i}\ .\label{GDRecursion}
\end{align} 
An obvious and anticlimactic interpretation of \eqref{GDBurgers}  is that the gravity dual of the $T\bar{T}$-deformed theory is BTZ black holes without a finite cutoff but instead with the deformed ADM mass
\be\label{deformedADMmass}
M_{\rm ADM}(R; \mu)={\pi R\over \mu}\left[1-\sqrt{1-{2\mu M_{\rm ADM}\over \pi R}}\right]\ .
\ee 
The recursion equation \eqref{GDRecursion} provides 
a more refined interpretation: At the $i$-th order of deformation, the dual geometry is the BTZ black holes \eqref{nonrotateBTZ} with the $S^1$ radius and the mass parameter
\begin{align}
 R^{(i)}=R\,\biggl(1-{\mu\over 2\pi R}M_{\rm ADM}^{(i-1)}\left(R;\mu\right)\biggr)\qquad\mbox{and}\qquad M^{(i)}
 =M\biggl({R\over R^{(i)}}\biggr)^2+{\cal O}(\mu^{i+1})\ .\label{GDatithorder}
\end{align}
In the continuum limit $i\to\infty$, these become
\begin{align}\label{GDatallorders}
R(\mu)={R\over 2}\left(1+\sqrt{1-{2\mu M\over\pi}}\right)\qquad\mbox{and}\qquad
M(\mu)=M\left({R\over R(\mu)}\right)^2\ .
\end{align}
We note that these yield the correct ADM mass $M_{\rm ADM}(R;\mu)=R(\mu)M(\mu)$ as it should. We will perform a consistency check of the dictionary \eqref{GDatallorders} in Section \ref{Sec:relation}.
Either way, in the random geometry approach, there is no finite cutoff unlike in the proposal of \cite{McGough:2016lol}, but there might be a way to relate the two viewpoints along the line of the idea advocated in \cite{Heemskerk:2010hk}.


\subsection{Thermal free energy}
\label{Sec:TE}

The thermal free energy was computed by Cardy in his original paper  \cite{Cardy:2018sdv}. Here we provide a more direct and simpler method to compute the thermal free energy so that the gravity dual becomes a straightforward adaptation of the field theory computation. We work in  Euclidean space and identify the circumference $2\pi R$ of the $S^1$ with  inverse temperature $\beta$, i.e.~$R=\beta/(2\pi)$.  We take the thermodynamic limit $\beta\ll L$ in which the thermal free energy is extensive and proportional to $L$.


\subsubsection{Field theory}
\label{subsecTE:FT}

For the computation of the thermal free energy $F(\beta)$ per unit length, the object of interest is the thermodynamic limit of the partition function
\be\label{PF}
\lim_{L\to\infty}Z(L,\beta)=\exp\left(-L\beta F(\beta)\right)=\exp\left(-L E_0(\beta)\right)\ ,
\ee
where $E_0(\beta)$ is the ground state energy in the crossed channel on a circle of circumference $\beta$. In the CFT case, the free energy is given by 
\be
\beta F(\beta)=E_0(\beta)=-{\pi c\over 6\beta}\ .
\ee
Applying the formula \eqref{deformcylinder}, the free energy of the ${\cal T}[\delta\mu]$ theory can be computed as
\begin{align}\label{TE1storder}
\hspace{-.2cm}
Z(L, \beta; \delta\mu)&= {\cal N}^{-1}\!\!\int dh \,e^{-L{\beta\over 4\delta\mu}h_{11}h_{22}}Z\left(\sqrt{1+h_{11}}L, \sqrt{1+h_{22}}\beta\right)
\equiv {\cal N}^{-1}\!\!\int dh\, e^{-L\beta {\cal F}[h]}.
\end{align}
In the thermodynamic limit $L\to\infty$, the deformed free energy is defined by
\be
\lim_{L\to\infty}Z(L, \beta; \delta\mu)=\exp\left(-L\beta F(\beta;\delta\mu)\right)\ .
\ee
Now, the free energy ${\cal F}[h]$ for a fixed $h$ in \eqref{TE1storder} is given by
\begin{align}\label{TE1storderv2}
\beta{\cal F}[h]&={\beta\over 4\delta\mu}h_{11}h_{22}+\sqrt{1+h_{11}}E_0\left(\sqrt{1+h_{22}}\beta\right)\nn\\
&=-{\pi c\over 6\beta}+{\beta\over 4\delta\mu}h_{11}h_{22}-\half(h_{11}-h_{22}){\pi c\over 6\beta}+{\cal O}(\delta\mu^2)
\end{align}
and the saddle point is at
\be
h^{\ast}_{11}=-h^{\ast}_{22}=-{\pi c\over 3\beta^2}\delta\mu\,\ .
\ee
Thus the free energy of the ${\cal T}[\delta\mu]$ theory is found to be
\begin{align}
F(\beta;\delta\mu)={\cal F}[h^{\ast}]=-{\pi c\over 6\beta^2}+{\pi^2c^2\over 36\beta^4}\delta\mu\ .
\end{align}
This reproduces the first-order correction in the all-order thermal free energy \cite{Cardy:2018sdv}
\begin{align}\label{freeEallorder}
F(\beta; \mu)={1\over 2\mu}\left[1-\sqrt{1+{2\pi c\mu \over 3\beta^2}}\right]\ .
\end{align}
Again, to be complete, the fluctuations about the saddle point must be taken into account. However, it is easy to see that the Gaussian integration over the fluctuations only yields a $L$-independent factor and thus does not contribute in the thermodynamic limit. 


\paragraph{$\bullet$ All orders I -- PDE (Burgers' equation)}

Applying the above procedure \eqref{TE1storderv2} for an infinitesimal deformation from the ${\cal T}[\mu]$ theory to the ${\cal T}[\mu+\delta\mu]$ theory, we find
\begin{align}
\beta F(\beta;\mu+\delta\mu)={\beta\over 4\delta\mu}h^{\ast}_{11}h^{\ast}_{22}
 +\half h^{\ast}_{11}\beta F(\beta;\mu)
+\half h^{\ast}_{22}\beta \del_{\beta}(\beta F(\beta;\mu))
 +\beta F(\beta;\mu)
 \label{kvsv11Mar20}
\end{align}
up to ${\cal O}(\delta\mu^2)$ terms,
where the saddle point is at
\be
h^{\ast}_{11}=-2\delta\mu\,\del_{\beta}\left(\beta F(\beta;\mu)\right)\ ,\qquad\qquad
h^{\ast}_{22}=-2\delta\mu\, F(\beta;\mu)\ .
\label{h_saddle_values}
\ee
This yields, again, the inviscid Burgers equation
\begin{align}\label{TEBurgers}
\del_{\mu}(\beta F(\beta;\mu))=-(\beta F(\beta;\mu))\,\del_{\beta}(\beta F(\beta;\mu))\ .
\end{align}
The solution with the initial condition $\beta F(\beta; 0)=-\pi
c/(6\beta)$ is indeed given by \eqref{freeEallorder}.

We can understand this in a different way.  Upon plugging in the saddle
point values~\eqref{h_saddle_values}, we see that the first two terms in
\eqref{kvsv11Mar20} cancel each other.  Therefore,
\begin{align}
\beta F(\beta;\mu+\delta\mu)
 =
 \beta F(\beta;\mu)
+\half h^{\ast}_{22}\beta \del_{\beta}(\beta F(\beta;\mu))
 =
 [\beta F(\beta;\mu)]|_{\beta \to \sqrt{1+h^*_{22}}\,\beta}.\label{ndqd11Mar20}
\end{align}
Namely, the evolution of $\beta F(\beta;\mu)$ can be interpreted as
coming from the repeated rescalings of~$\beta$.  Let us describe this
evolution by $\beta(\mu)$; namely, at $\mu=0$ we have $\beta(0)=\beta$,
which evolves to $\beta(\mu)$ at $\mu$.  Note that $\beta(\mu)$ is a
function of the coupling $\mu$ as well as the initial condition $\beta$.
The relation \eqref{ndqd11Mar20} says that, as we change
$\mu\to\mu+\delta\mu$, the change in $\beta(\mu)$ is accounted for by
the rescaling $\beta\to \sqrt{1+h^*_{22}}\,\beta$.  Therefore,
\begin{align}
 \delta \beta(\mu) ={\partial \beta(\mu)\over \partial \beta}{h^*_{22}\beta\over 2}.
\end{align}
Using the saddle point value  of $h^*_{22}$ in 
\eqref{h_saddle_values}, we obtain the evolution equation for $\beta(\mu)$:
\begin{align}
 {\partial \beta(\mu)\over \partial \mu}
 =-{\partial \beta(\mu)\over \partial \beta}\,\beta F(\beta;\mu).
\end{align}
Given the expression for $F(\beta;\mu)$ in \eqref{freeEallorder}, it is
not difficult to solve this with the initial condition $\beta(0)=\beta$.
The solution is
\begin{align}
 \beta(\mu)=
 \frac{\beta}{2}  \left(1+\sqrt{1+\frac{2 \pi  c \mu }{3 \beta^2}}\right).\label{mplj11Mar20}
\end{align}
We can readily check that the partition function of the $\mu$ deformed
theory, $F(\beta;\mu)$, is equal to the partition function of the
undeformed theory, i.e.~CFT, at inverse temperature
$\beta=\beta(\mu)$. Namely,
\begin{align}
 F(\beta;\mu)=F(\beta(\mu),0).
\end{align}


\paragraph{$\bullet$ All orders II -- order-by-order iteration}

In a similar way to the case of the energy spectrum, we can find the free energy by iteration order by order in $\mu$ by replacing
\be
R^{(i)}\to \beta^{(i)}/(2\pi)\qquad\quad\mbox{and}\qquad\quad E^{(i)}_n(R;\mu)\to \beta^{(i)}F^{(i)}(\beta;\mu)\equiv{\cal F}^{(i)}(\beta;\mu)\ .
\ee
Note that this, in particular, implies that at the $i$-th order
\be\label{ithbeta}
{\cal F}^{(i)}(\beta;\mu)=-{\pi c\over 6\beta^{(i)}}+{\cal O}(\mu^{i+1})\qquad\mbox{with}\qquad
\beta^{(i)}=\beta\left(1+\half h^{\ast(i)}_{22}\right)\simeq \beta\sqrt{1+h^{\ast(i)}_{22}}
\ee
with $h_{22}^{\ast(i)}=-(2\mu/\beta){\cal F}^{(i-1)}(\beta;\mu)$. 
Then, with these replacements, we obtain from \eqref{recursion}  the recursion equation 
\begin{align}\label{TErecursion}
{\cal F}^{(i)}(\beta;\mu)=E_0(\beta)+\left[{\mu\over \beta}\left({\cal F}^{(i-1)}(\beta;\mu)\right)^2\right]_{i}
\end{align}
whose continuum limit can be solved by the all-order thermal free energy \eqref{freeEallorder}.


\subsubsection{Gravity dual}
\label{subsecTE:GD}

The gravity dual starts with the Euclidean BTZ black holes:
\begin{align}\label{EBTZ}
ds_{\rm EBTZ}^2={1\over 4}\left(\rho-\frac{8GM}{\rho}\right)^2d\tau^2+{1\over 4}\left(\rho+\frac{8GM}{\rho}\right)^2dy^2+{d\rho^2\over\rho^2}\ ,
\end{align}
where $\tau\sim\tau+\beta$ and $0\le y \le L$ and we reinstated the 3$d$ Newton constant $G$. 
As usual, the smoothness at the horizon $\rho=\sqrt{8GM}$ requires $\beta=2\pi/\sqrt{8GM}$.
Note that in comparison to the field theory, this description corresponds to the crossed channel. A key observation is that under the rescaling 
$\tau\mapsto\sqrt{1+h_{11}}\tau$, the absence of the conical singularity requires that
\be\label{conesingfree}
M\mapsto {M\over 1+h_{11}}\ .
\ee
Although the arguments that lead to it are different, this is the thermal-counterpart of the statement \eqref{BTZdeformation} in the case of the energy spectrum.

\medskip
The gravity dual of the partition function \eqref{PF} at large $c$ is simply the classical gravity partition function evaluated on the Euclidean BTZ black holes.
A detail of the computation is provided in Appendix \ref{Appendix:Onshellaction}\@. With the deformation \eqref{conesingfree} taken into account, the renormalized on-shell action is given by
\begin{align}\label{EBTZaction}
S_{\rm EBTZ}\left(g^{\rm bdy}_{ii}=1+h_{ii}\right)=-L \beta{M\over 2\pi}\sqrt{1+h_{22}\over 1+h_{11}}=-L \beta{\pi c\over 6\beta^2}\sqrt{1+h_{22}\over 1+h_{11}}\ ,
\end{align}
where we used $M=(2\pi/\beta)^2/(8G)=(2\pi/\beta)^2(c/12)$.
The gravity dual of the infinitesimal deformation \eqref{TE1storder} is then given by
\begin{align}\label{GDTE1storder}
\hspace{-.2cm}
e^{-S_{\rm EBTZ}(L,\beta; \delta\mu)}&= {\cal N}^{-1}\!\!\int dh \,e^{-L{\beta\over 4\delta\mu}h_{11}h_{22}}e^{-S_{\rm EBTZ}\left(g^{\rm bdy}_{ii}=1+h_{ii}\right)}\ ,
\end{align}
where $S_{\rm EBTZ}(L,\beta; \delta\mu)=L\beta F(L, \beta; \delta\mu)$.
Since the deformed on-shell action \eqref{EBTZaction} for a fixed $h$ is identical to that of the CFT dual in \eqref{TE1storderv2} by exchanging $h_{11}\leftrightarrow h_{22}$, we are guaranteed to obtain the same results as in \eqref{TEBurgers} and \eqref{TErecursion}.

\medskip
In addition to the dictionary \eqref{deformedADMmass} -- \eqref{GDatallorders} derived in the energy spectrum analysis, 
a new piece of data to the gravity dual, inferred from \eqref{ithbeta}, is the inverse Hawking temperature at the $i$-th order
\begin{align}
\beta^{(i)}=\beta\left(1-{\mu\over\beta}{\cal F}^{(i-1)}(\beta;\mu)\right)
\end{align}
which, in the continuum limit $i\to\infty$, becomes
\be\label{resizingbeta}
\beta(\mu)={\beta\over 2}\left(1+\sqrt{1+{\pi \mu \over G\beta^2}}\right)\ .
\ee
This is the same as \eqref{mplj11Mar20}.
Note that the consistency of \eqref{resizingbeta} and \eqref{GDatallorders} requires a relation between $\beta$ and $M$
\be
\beta=2\pi\sqrt{\frac{1}{8GM}\left(1  -{2  \mu  M\over\pi}\right)}\ ,
\ee
which we will elaborate  in the next section.


\subsection{Micro-canonical to canonical ensemble}
\label{Sec:relation}

Here we provide another perspective on the energy spectrum in Section \ref{Sec:ES} and the thermal free energy in Section \ref{Sec:TE}.
Namely, they must be related by
\begin{align}
e^{-L\beta F(\beta;\mu)}=\int dM\, e^{S(M) - \beta E(M;\mu)}\qquad\mbox{with}\qquad S(M)= L\sqrt{cM\over 3}\ ,
\end{align}
where the entropy $S(M)$ is Cardy's formula in the field theory and the BTZ black hole entropy in gravity.
Indeed, for the energy spectrum \eqref{SZformula}
\be
E(M;\mu)={L\over 2\mu}\left[1-\sqrt{1-{2\mu M\over \pi}}\right]\ ,
\ee
the $M$-integral has the saddle point
\be\label{Msaddle}
M_{\ast}= \frac{\pi ^2 c}{3 \beta ^2+2 \pi  c \mu }\qquad\Longleftrightarrow\qquad
\beta=\pm\sqrt{\frac{\pi^2 c }{3M_{\ast}}\left(1  -{2  \mu  M_{\ast}\over\pi}\right)}\ ,
\ee
and the saddle point approximation yields the thermal free energy of the ${\cal T}[\mu]$ theory \eqref{freeEallorder}
\begin{align}
\beta F(\beta;\mu)=-S(M_{\ast}) + \beta E(M_{\ast})={\beta\over 2\mu}\left[1-\sqrt{1+{2\pi c\mu \over 3\beta^2}}\right]\ .
\end{align}
Note that without deformation, the saddle point $M_{\ast}$ obviously obeys the relation between the ``mass'' and the inverse Hawking temperature $\beta=2\pi/\sqrt{8GM_{\ast}}$ with $c=3/(2G)$. In fact, even though it may not be obvious, the same holds true in the deformed case: From the dictionary  \eqref{GDatallorders} derived from the energy spectrum analysis, we can compute the inverse Hawking temperature of the ${\cal T}[\mu]$ theory, $\beta(\mu)=2\pi/\sqrt{8GM(\mu)}$. Meanwhile, from the dictionary \eqref{resizingbeta} derived from the free energy analysis, we found the resizing of the thermal circle. These two must agree, i.e.
\begin{align}
\beta(\mu)={\pi\over \sqrt{8GM}}\left(1+\sqrt{1-{2\mu M\over\pi}}\right)={\beta\over 2}\left(1+\sqrt{1+{\pi \mu \over G\beta^2}}\right)\ .
\end{align}   
Indeed, solving this equation for $\beta$ yields the saddle point relation \eqref{Msaddle}
\be
\beta=2\pi\sqrt{\frac{1}{8GM}\left(1  -{2  \mu  M\over\pi}\right)}\ .
\ee
This provides a consistency check of our proposed gravity dual.

\medskip
A few more remarks are in order: An unconventional property is that for a $\mu>0$ deformation, the ``energy'' $M_{\ast}$ reaches a maximum value $M_{\rm max}=\pi/(2\mu)$ and thus remains finite in the infinite temperature limit, $\beta\to 0$. In contrast, for $\mu<0$ corresponding to the ``Hagedorn  branch,''  the saddle point \eqref{Msaddle} and the free energy both indeed indicate that there is a maximum limiting temperature $\beta_{\rm Hag}^{-1}=\sqrt{3/(2\pi|\mu| c)}$.


\section{Correlation functions on $\mathbb{R}^2$}
\label{Sec:Correlators}

In Cardy's original work \cite{Cardy:2018sdv}, the random geometry approach was not applied to correlation functions.\footnote{Aharony and Vaknin utilized this approach to the computation of stress tensor correlators \cite{Aharony:2018vux}.} 
Here we generalize the idea reviewed in Section \ref{Sec:RandomGeometry} to the case when singularities are present due to local operator insertions. Our new method reproduces the all-order correlation functions computed by Cardy in a different way \cite{Cardy:2019qao} and renders the holographic adaptation straightforward. 

\medskip
For clarity and convenience, we quote the Hubbard-Stratonovich transformation of an infinitesimal $T\bar{T}$ deformation in Section \ref{Sec:RandomGeometry}:
\begin{align}\label{Correlators:InfinitesimalDeform}
\exp\left(-\delta S\right)\propto \int [dh]\exp\biggl[-{1\over 8\delta\mu}\int d^2x\,\epsilon^{ik}\epsilon^{jl}h_{ij}h_{kl}+\int d^2x\, h_{ij}T^{ij}\biggr] 
\end{align}
which is dominated by the saddle point
\begin{align}\label{Correlator:saddlegen}
{1\over 4\delta\mu}h_{22}^{\ast}=T_{11}\ , \qquad {1\over 4\delta\mu}h_{11}^{\ast}=T_{22}\ ,\qquad
{1\over 4\delta\mu}h_{12}^{\ast}=-T_{12}\ ,\qquad {1\over 4\delta\mu}h_{21}^{\ast}=-T_{21}\ .
\end{align}

To this point, everything is exactly the same as in Section \ref{Sec:RandomGeometry}. Then if the argument in Section~\ref{Sec:RandomGeometry} remains intact, we would have to conclude that there is no effect of the $T\bar{T}$ deformation on $\mathbb{R}^2$. 
As remarked in footnote \ref{singularities}, however, when local operators are inserted at $x_a$ ($a=1,\dots,n$), they create singularities and source the stress-energy, 
\begin{align}\label{conservationwithsource}
\del_iT^{ij}(x)=\sum_{a=1}^n J^j(x_a, \del_{x_a})\delta^2(x-x_a)\ .
\end{align}
This is where the argument in Section \ref{Sec:RandomGeometry} needs to
be corrected in order to compute correlation functions.  The precise
form of the source $J^j(x_a, \del_{x_a})$ is determined by the
Ward-Takahashi (WT) identity \cite{Cardy:2019qao}\footnote{A factor of
$-1/2$ on the RHS is due to our normalization of the stress tensor
\eqref{Tij_def}.   The more standard definition of the stress tensor  $T^{ij}_{\rm std}={2\over\sqrt{g}}{\delta S\over \delta g_{ij}}$ is related 
to ours by $T^{ij}_{\rm here}=-\half T^{ij}_{\rm std}$.\label{Tnormalization}}
\begin{align}
(2\pi)T_{ij}(x)\prod_{a=1}^n{\cal O}_a(x_a)
 &=
 -\half\sum_{a=1}^n\Biggl[
 \del_{x^i}\ln{|x-x_a|\over\varepsilon}\del_{x^j_a}
 +\del_{x^j}\ln{|x-x_a|\over\varepsilon}\del_{x^i_a}
 -\delta_{ij}\del_{x^k}\ln{|x-x_a|\over\varepsilon}\del_{x^k_a}
 \nn\\
&\quad
 -\left(\Delta_a+\gamma_a(\mu)\right)\del_i\del_j\ln{|x-x_a|\over\varepsilon}+\pi\delta_{ij}\left(\Delta_a+\gamma_a(\mu)\right)\delta^2(x-x_a)\nn\\
&\quad
 +\pi\delta_{ij}\gamma_a(\mu)\delta^2(x-x_a)\Biggr]\prod_{b=1}^n{\cal O}_b(x_b)\ ,\label{TijWTid}
\end{align}
where $\varepsilon$ is a UV cutoff, $\Delta_a$ is the dimension of the operator ${\cal O}_a$, and $\gamma_a(\mu)$ is the ``anomalous dimension'' of ${\cal O}_a$, as will be elaborated. $\gamma_a(\mu)$ depends on all the positions $x_1,\dots,x_n$ of the insertions, although the dependence is not explicitly shown. It vanishes for CFT, i.e.\ $\gamma_a(0)=0$.
The second and third lines may require more explanations and we will argue that this is the correct form as we proceed.  
At any rate, this implies the conservation law~\eqref{conservationwithsource} in the integrated form
\begin{align}\label{TijWTidv2}
T_{ij}(x)&=\sum_{a=1}^n\Biggl[-{1\over 4\pi}\biggl(
\del_{x^i}\ln{|x-x_a|\over\varepsilon}\del_{x^j_a}
+ \del_{x^j}\ln{|x-x_a|\over\varepsilon}\del_{x^i_a}
-\delta_{ij}\del_{x^k}\ln{|x-x_a|\over\varepsilon}\del_{x^k_a}\biggr)\!\ln\left\langle\prod_{b=1}^n{\cal O}_b(x_b)\right\rangle\nn\\
&\hspace{1.3cm}+{1\over 4\pi}\left(\Delta_a+\gamma_a(\mu)\right)\del_i\del_j\ln{|x-x_a|\over\varepsilon}-{1\over 4}\delta_{ij}\left(\Delta_a+\gamma_a(\mu)\right)\delta^2(x-x_a)\nn\\
&\hspace{1.3cm}-{1\over 4}\delta_{ij}\gamma_a(\mu)\delta^2(x-x_a)\Biggr].
\end{align}

Here we give some justifications of the form of the WT identity
\eqref{TijWTid}.  First, we note that taking a derivative of it yields a more
familiar form of the WT identity \cite{Fujikawa:1980rc}
\begin{align}
\hspace{-.3cm}
\del_iT_{ij}(x)\prod_{a=1}^n{\cal O}_a(x_a)=-\half\sum_{a=1}^n\biggl[\delta^2(x-x_a)\del_{x^j_a}
-{\Delta_a\over 2}\del_{x^j}\delta^2(x-x_a)\biggr]\prod_{b=1}^n{\cal O}_b(x_b)\ .
\end{align}
Next, multiplying this by $x^j$ and integrating over $x$ we get
\begin{align}
\int d^2x\,T_{i}^i(x)\prod_{a=1}^n{\cal O}_a(x_a)=\half\sum_{a=1}^n\biggl[x_a^i\del_{x^i_a}
+\Delta_a\biggr]\prod_{b=1}^n{\cal O}_b(x_b)\ .\label{kjao12Mar20}
\end{align}
Meanwhile, taking a trace of \eqref{TijWTid} and integrating over $x$ yields
\begin{align}
\int d^2x\,T_{i}^i(x)\prod_{a=1}^n{\cal O}_a(x_a)=-\half\sum_{a=1}^n\gamma_a(\mu)\prod_{b=1}^n{\cal O}_b(x_b)\ .\label{kjcg12Mar20}
\end{align}
By combining \eqref{kjao12Mar20} and \eqref{kjcg12Mar20}, we find that
\be
\sum_{a=1}^n\biggl[x_a^i\del_{x^i_a}+\Delta_a+\gamma_a(\mu)\biggr]\prod_{b=1}^n{\cal O}_b(x_b)=0\ ,
\label{lbia12Mar20}
\ee
which generalizes the relation for CFT ($\mu=0$) for which $\gamma_a=0$.
Furthermore, as we will see below, the consistency of the WT identity with the flow equation~\cite{Smirnov:2016lqw, Cavaglia:2016oda} fixes the form of the ``anomalous dimension'' $\gamma_a(\mu)$ and, in particular, we will find the relation
\be\label{muscaleviolation}
\sum_{a=1}^n\gamma_a(\mu)\prod_{b=1}^n{\cal O}_b(x_b)=\left[2\mu{\del\over\del\mu}+\varepsilon{\del\over\del\varepsilon}\right]\prod_{a=1}^n{\cal O}_a(x_a)
\ee
which reflects the fact that the coupling $\mu$ and the cutoff $\varepsilon$ have length dimension 2 and 1, respectively, and are the only sources of scale symmetry violation in the $T\bar{T}$-deformed CFTs.
Combined with \eqref{lbia12Mar20} this then yields the WT identity
\begin{align}
 \label{CSequation}
 \biggl[\sum_{a=1}^n\left(x_a^i\del_{x^i_a}+\Delta_a\right)+2\mu{\del\over\del\mu}+\varepsilon{\del\over\del\varepsilon}\biggr]\prod_{b=1}^n{\cal O}_b(x_b)=0
\end{align}
which is a form of the Callan-Symanzik equation discussed in \cite{Cardy:2019qao}.

\medskip
Now, the 2$d$ metric can generally be decomposed into the diffeomorphism and Weyl parts:
\be\label{Correlators:metric}
h_{ij}=\del_i\alpha_j+\del_j\alpha_i+\delta_{ij}\Phi\ .
\ee
We take $\alpha_i$ to be single-valued.
We further find it most convenient to split the Weyl factor~$\Phi$ into
\be\label{Weylsplit}
\Phi = -\del_k\alpha_k + \phi
\ee
so that $\phi$ alone represents the trace part of the saddle point metric and stress tensor.
Then the saddle point equation \eqref{Correlator:saddlegen} can be solved by
\begin{align}
\alpha_i(x)&={\delta\mu\over \pi}\sum_{a=1}^n\ln{|x-x_a|\over\varepsilon}\del_{x^i_a}\ln\left\langle\prod_{b=1}^n{\cal O}_b(x_b)\right\rangle
+{\delta\mu\over 2\pi}\sum_{a=1}^n\left(\Delta_a+\gamma_a(\mu)\right){(x-x_a)_i\over |x-x_a|^2},
\label{Correlators:alpha}\\
\phi(x)&=-\delta\mu\sum_{a=1}^n\gamma_a(\mu)\delta^2(x-x_a)\ .\label{Correlators:Weyl}
\end{align}
We note that this is our choice of ``gauge'' and the solution is not unique because there is an arbitrariness in the way we split the metric into the diffeomorphism and Weyl parts.


\paragraph{$\bullet$ An alternative and more intuitive method:}

There is an alternative and more intuitive way to incorporate the local operator singularities. Instead of using the conservation law and the WT identity, we can simply exponentiate correlation functions to find the saddle point metric.
As in the previous sections, the $hT$ term can be interpreted as the change in the metric $g_{ij}\mapsto g_{ij}+h_{ij}$. Under infinitesimal coordinate transformations $x_i\mapsto x_i+\alpha_i$, the correlation functions transform as 
\begin{align}
 \label{Correlatortransform}
 \left\langle\prod_{a=1}^n{\cal O}_a(x_a)\right\rangle\qquad\stackrel{\rm Diff}{\xrightarrow{\hspace*{0.8cm}}}\qquad
 \Biggl\langle\prod_{a=1}^n\left(1+\del_k\alpha_k(x_a)\right)^{\Delta_a\over 2} {\cal O}_a(x_a+\alpha(x_a))\Biggl\rangle\ .
\end{align}
In addition, there should be ``wavefunction renormalization'' due to ``anomalous'' Weyl scalings
\be\label{CorrelatortransformWeyl}
\hspace{-.2cm}
\left\langle\prod_{a=1}^n{\cal O}_a(x_a)\right\rangle\,\stackrel{{\rm Diff}+{\rm Weyl}}{\xrightarrow{\hspace*{1.2cm}}}\,
\left\langle\prod_{a=1}^n(1+\Phi(x_a))e^{-{\gamma_a\over 2}}\left(1+\del_k\alpha_k(x_a)\right)^{\Delta_a\over 2} {\cal O}_a(x_a+\alpha(x_a))\right\rangle,
\ee
where $\Phi=\phi-\del_k\alpha_k$ and we abbreviated $\gamma_a(\mu)$ as
$\gamma_a$.  For CFT $(\mu=0)$, $\gamma_a=0$ and there is
no anomalous rescaling.
Thus an infinitesimal
deformation of correlation functions can be computed as
\begin{align}\label{infinitesimaldeformedcorrelators}
\hspace{-.2cm}
\left\langle\prod_{a=1}^n{\cal O}_a(x_a)\right\rangle_{\mu+\delta\mu}
\hspace{-.4cm}=&\,\, {\cal N}^{-1}\int d\alpha\exp\Biggl[-{1\over 4\delta\mu}\int d^2x\left(\alpha_i\Box\alpha_i+\phi^2\right)
+\ln\left\langle\prod_{a=1}^n{\cal O}_a(x_a+\alpha(x_a))\right\rangle\nn\\
&\hspace{2.9cm}+\sum_{a=1}^n\biggl({\Delta_a+\gamma_a\over 2}\del_k\alpha_k(x_a)-{\gamma_a\over 2}\phi(x_a)\biggr)\Biggr]
\end{align}
where we used that for the metric of the form \eqref{Correlators:metric} with \eqref{Weylsplit} the Gaussian $h$ action becomes, for single-valued $\alpha_i$,
\be
\int d^2x\,\epsilon^{ik}\epsilon^{jl}h_{ij}h_{kl}=-\int d^2x\, h_{ij}h^{ij}=2\int d^2x\left(\alpha_i\Box\alpha_i+\phi^2\right)\ .
\ee
The exponentiated correlators in \eqref{infinitesimaldeformedcorrelators} can be expanded to the first order in $\alpha$ as
\begin{align}
\ln\left\langle\prod_{a=1}^n{\cal O}_a(x_a+\alpha_a)\right\rangle
=\ln\left\langle\prod_{a=1}^n{\cal O}_a(x_a)\right\rangle
+\sum_{a=1}^n\alpha^i(x_a){\del\over\del x_a^i}\ln\left\langle\prod_{b=1}^n{\cal O}_b(x_b)\right\rangle.
\end{align}
Thus the saddle point equation reads
\begin{align}
\Box\alpha^{\ast}_i(x)&=2\delta\mu\sum_{a=1}^n\delta^2(x-x_a){\del\over\del x_a^i}\ln\left\langle\prod_{b=1}^n{\cal O}_b(x_b)\right\rangle
+\delta\mu\sum_{a=1}^n\left(\Delta_a+\gamma_a\right)\del_i\delta^2(x-x_a)\ ,\label{Correlators:saddleeqn}\\
\phi^{\ast}(x)&=-\delta\mu\sum_{a=1}^n\gamma_a\delta^2(x-x_a)\ .\label{Weylsaddle}
\end{align}
The solution to the first equation precisely agrees with \eqref{Correlators:alpha} obtained by using the conservation law and the WT identity.
We then find at the saddle point that \eqref{infinitesimaldeformedcorrelators} yields
\begin{align}\label{Correlators:saddleaction}
e^{-\delta S_{\rm saddle}}
 &=\left\langle\prod_{a=1}^n{\cal O}_a(x_a)\right\rangle e^{{1\over 4\delta\mu}
\int d^2x\left(\alpha^{\ast}_i(x)\Box\alpha^{\ast}_i(x)+\phi^{\ast}(x)^2\right)}
\notag\\
&=\left\langle\prod_{a=1}^n{\cal O}_a(x_a)\right\rangle
\exp\Biggl[{\delta\mu\over 2\pi}\sum_{a\ne b}\ln{|x_a-x_b|\over\varepsilon}\,
{\del\over\del x_a^i}\ln\left\langle\prod_{c=1}^n{\cal O}_c(x_c)\right\rangle
{\del\over\del x_b^i}\ln\left\langle\prod_{d=1}^n{\cal O}_d(x_d)\right\rangle\nn\\
&\hspace{1.2cm}-{\delta\mu\over 2\pi}\sum_{a\ne b}\left(\delta_a+\gamma_a(\mu)\right){x_a^i-x_b^i\over |x_a-x_b|^2}{\del\over\del x_b^i}\ln\left\langle\prod_{c=1}^n{\cal O}_c(x_c)\right\rangle
\Biggr]
\end{align}
where we dropped divergent terms of the type $\ln 0$, $1/0$ and $\delta^2(0)$. We assume that these divergences can be absorbed into local renormalization of fields.


\subsection{First-order correction}
\label{Sec:FO}

As an illustration of this method, we first discuss the first-order correction to two point functions. The undeformed two point functions are of the form
\begin{align}
\langle{\cal O}(x_1){\cal O}(x_2)\rangle_0={1\over |x_1-x_2|^{2\Delta}}\ 
\end{align}
where $\Delta$ is the dimension of $\cal O$.
Applying the above formulas \eqref{infinitesimaldeformedcorrelators} and \eqref{Correlators:saddleaction}, we obtain
\begin{align}
\langle{\cal O}(x_1){\cal O}(x_2)\rangle_{\delta\mu}
&={1\over |x_1-x_2|^{2\Delta}}\exp\biggl(-{4\delta\mu\,\Delta^2\over \pi}{\ln(|x_1-x_2|/\widetilde{\varepsilon})\over |x_1-x_2|^{2}}\biggr)\nn\\
&={1\over |x_1-x_2|^{2\Delta}}\biggl[1-{4\delta\mu\,\Delta^2\over \pi}{\ln(|x_1-x_2|/\widetilde{\varepsilon})\over |x_1-x_2|^{2}}\biggr]+{\cal O}(\delta\mu^2)\ ,
\label{lzmh12Mar20}
\end{align}
where we absorbed the non-logarithmic power correction into the UV cutoff $\widetilde{\varepsilon}$.
This agrees with the first-order result in \cite{Kraus:2018xrn} and \cite{Cardy:2019qao}.\footnote{In these two papers, the $T\bar{T}$-deformation coupling is denoted by $\lambda$. Our coupling $\mu$ is related to their $\lambda$ by $\delta\lambda^{\rm KLM}=-\delta\mu/\pi^2$ and $\delta\lambda^{\rm Cardy}=-\delta\mu/(4\pi)$, respectively. Our first-order two point functions agree with those in \cite{Kraus:2018xrn} but differ by a factor of 2 from those in \cite{Cardy:2019qao}.}
To be complete, the fluctuations about the saddle point have to be taken into account. As we will discuss in the next subsection, the fluctuations make no contribution to the first order. However, at higher orders, this is not the case anymore and the fluctuations are an important part of higher order corrections.


\subsection{All-order corrections}
\label{Sec:AO}

We apply the above method to an infinitesimal deformation from the ${\cal T}[\mu]$ theory to the ${\cal T}[\mu+\delta\mu]$ theory. 
As is the case in  \cite{Cardy:2019qao}, we now focus on leading logarithmic corrections. This means that we can neglect the second term on the RHS of the saddle point equation~\eqref{Correlators:saddleeqn} and the Weyl factor \eqref{Weylsaddle}. In other words, we ignore the term proportional to $\Delta_a\del_i\delta^2(x-x_a)$ in the WT identity of $\del_iT^{ij}$, or equivalently, the Jacobian and Weyl factors in the transformation of correlators \eqref{CorrelatortransformWeyl}. As commented above, the Gaussian fluctuations about the saddle point makes a nonvanishing contribution to higher order corrections in $\mu$. Expanding the action about the saddle point, $\alpha_i=\alpha_i^{\ast}+\delta\alpha_i$, to the quadratic order in fluctuations $\delta\alpha_i$, one finds that
\begin{align}
-\delta S=&\,{\delta\mu\over 2\pi}\sum_{a\neq b}\ln{|x_a-x_b|\over\varepsilon}
{\del\over\del x_a^i}\ln\left\langle\prod_{c=1}^n{\cal O}_c(x_c)\right\rangle_{\mu}
{\del\over\del x_b^i}\ln\left\langle\prod_{d=1}^n{\cal O}_d(x_d)\right\rangle_{\mu}\nn\\
&-{1\over 4\delta\mu}\int d^2x\,\delta\alpha_i\Box\delta\alpha_i+\half\sum_{a,b}\delta\alpha^i(x_a)\delta\alpha^j(x_b){\del^2\over\del x_a^i\del x_b^j}
\ln\left\langle\prod_{c=1}^n{\cal O}_c(x_c)\right\rangle_{\mu}\ .
\end{align}
Note that the fluctuation contribution in the second line vanishes in the case of two point functions with $\mu=0$. This justifies the absence of the fluctuation contribution in the previous subsection \ref{Sec:FO}.
Performing the Gaussian integrals over $\delta\alpha_i$ and taking into account the normalization factor ${\cal N}^{-1}$, we obtain
\begin{align}\label{Correlators:saddleaction_allo}
-(\delta S_{\rm saddle}+\delta S_{\rm fluctuation})
=&\,{\delta\mu\over 2\pi}\sum_{a\ne b}\ln{|x_a-x_b|\over\varepsilon}\,
{\partial\over\partial x_a^i}\ln\left\langle\prod_{c=1}^n{\cal O}_c(x_c)\right\rangle_{\mu}
{\partial\over\partial x_b^i}\ln\left\langle\prod_{d=1}^n{\cal O}_d(x_d)\right\rangle_{\mu}\nn\\
&+{\delta\mu\over 2\pi}\sum_{a\ne b}\ln{|x_a-x_b|\over\varepsilon}\,{\partial^2\over\partial x_a^i\partial x_b^i}
\ln\left\langle\prod_{c=1}^n{\cal O}_c(x_c)\right\rangle_{\mu}\ .
\end{align}
From \eqref{infinitesimaldeformedcorrelators}, this yields a PDE for the deformed correlation functions
\begin{align}\label{Correlators:PDE}
{\partial\over\partial\mu}\left\langle\prod_{a=1}^n{\cal O}_a(x_a)\right\rangle_{\mu}
={1\over 2\pi}\sum_{a\ne b}\ln{|x_a-x_b|\over\varepsilon}\,{\partial^2\over\partial x_a^i\partial x_b^i}\left\langle\prod_{c=1}^n{\cal O}_c(x_c)\right\rangle_{\mu}
\ .
\end{align}
This agrees with the result in \cite{Cardy:2019qao} derived in a different method. In particular, in the case of two point functions, this PDE can be solved, in momentum space, by
\begin{align}\label{2ptallorder}
\left\langle\tilde{\cal O}(k)\tilde{\cal O}(-k)\right\rangle_{\mu}
\propto
k^{2(\Delta-1)}
e^{-{\mu\over 2\pi} k^2\ln (k^2\varepsilon^2)}
\end{align} 
summing up leading logarithmic corrections to all orders in $\mu$. Note that this suggests how the operators can be renormalized  \cite{Cardy:2019qao},
\be
\tilde{\cal O}^{\rm ren}(k)=\varepsilon^{{\mu\over 2\pi}k^2}\tilde{\cal O}(k)
\ee
in terms of which correlators are finite as we remove the cutoff $\varepsilon\to 0$.

\medskip
Finally, as a consistency check, we now determine the form of the ``anomalous dimension''~$\gamma_a(\mu)$ from the flow equation \cite{Smirnov:2016lqw, Cavaglia:2016oda}
\be
\left[\int d^2x\, T_i^i+\half\,\varepsilon{\del\over\del\varepsilon}\right]\prod_{a=1}^n{\cal O}_a(x_a)=-4\mu\int d^2x\, \det T_{ij}\prod_{a=1}^n{\cal O}_a(x_a)\ ,
\ee
where we have taken into account the fact that in the presence of the cutoff $\varepsilon$, the coupling $\mu$ is not the only source of scale symmetry violation in the theory. As remarked in footnote~\ref{Tnormalization}, the convention of our stress tensor is $-1/2$ of the standard one.
From \eqref{TijWTidv2} the integral of the stress tensor trace on the LHS is given by
\be
\int d^2 x T_i^i=-\half\sum_{a=1}^n\gamma_a(\mu)\ .
\ee
Meanwhile, recalling the Hubbard-Stratonovich representation \eqref{Correlators:InfinitesimalDeform} of the $T\bar{T}$ operator, we see that \eqref{Correlators:saddleaction} is $+4\delta\mu\int d^2x\,\det T_{ij}$ by definition and we then find that
\be
-4\mu\int d^2x\, \det T_{ij}=-{\mu\over 2\pi}\sum_{a\ne b}\ln{|x_a-x_b|\over\varepsilon}\,
\frac{{\partial^2\over\partial x_a^i\partial x_b^i}\left\langle\prod_{c=1}^n{\cal O}_c(x_c)\right\rangle_{\mu}}{\left\langle\prod_{d=1}^n{\cal O}_d(x_d)\right\rangle_{\mu}}
\ee
for the leading logarithmic corrections. We thus obtain
\begin{align}
\sum_a \gamma_a(\mu)&={\mu\over \pi}\sum^n_{a\neq b}\ln{|x_a-x_b|\over\varepsilon}\,
\frac{{\partial^2\over\partial x_a^i\partial x_b^i}\left\langle\prod_{c=1}^n{\cal O}_c(x_c)\right\rangle_{\mu}}{\left\langle\prod_{d=1}^n{\cal O}_d(x_d)\right\rangle_{\mu}}
+\varepsilon{\del\over\del\varepsilon}\ln\left\langle\prod_{a=1}^n{\cal O}_a(x_a)\right\rangle_{\mu}\ .\label{lsdy12Mar20}
\end{align}
Finally, the PDE for correlators \eqref{Correlators:PDE} then implies that
\be
\sum_{a=1}^n\gamma_a(\mu)\prod_{b=1}^n{\cal O}_b(x_b)=\left[2\mu{\del\over\del\mu}+\varepsilon{\del\over\del\varepsilon}\right]\prod_{a=1}^n{\cal O}_a(x_a)
\ee
as alluded in \eqref{muscaleviolation}. In other words, the entire corrections to the correlators can be thought of as coming from anomalous dimensions that depend on the positions of the insertions, as may be suggested by the form of the two point functions \eqref{2ptallorder}.

For example, for the two point function for which $\Delta_1=\Delta_2=\Delta$ and
$\gamma_1=\gamma_2=\gamma$, the relation \eqref{lsdy12Mar20} gives
\begin{align}
\gamma(\mu)&={\mu\over \pi}\ln{|x_1-x_2|\over\varepsilon}\,
\frac{{\partial^2\over\partial x^i_1\partial x^i_2}\langle {\cal O}(x_1){\cal O}(x_2)\rangle_{\mu}}{\langle{\cal O}(x_1){\cal O}(x_2)\rangle_{\mu}}
+{1\over 2}\varepsilon{\del\over\del\varepsilon}\ln\langle{\cal O}(x_1){\cal O}(x_2)\rangle_{\mu}.
\end{align}
Using the first order expression
\eqref{lzmh12Mar20}, we find
\begin{align}
 \gamma(\mu)={2\Delta^2(1-2\ln{|x_1-x_2|\over\varepsilon})\over \pi |x_1-x_2|^2}\mu+\cO(\mu^2).
\end{align}


\subsection{Gravity dual}
\label{subsecCF:GD}

It is a matter of straightforward translation of languages to find the gravity dual description. Rephrasing the above field theory computation in the language of gravity via the GKP-W dictionary \cite{Gubser:1998bc, Witten:1998qj}, it reads
\begin{align}
\left\langle\prod_{a=1}^n{\cal O}_a(x_a)\right\rangle_{\delta\mu}=
{\cal N}^{-1}\int d\alpha{\delta^nZ_{{\rm AdS}_3}[\phi(\rho,x)\stackrel{\rho\to\infty}{\longrightarrow}J(x)\rho^{\Delta-2}] \over\delta J(x_1)\cdots\delta J(x_n)}\Biggr|_{x\to x+\alpha}\!\!\! e^{-{1\over 4\delta\mu}\int d^2x\,\alpha_i\Box\alpha_i},
\end{align}
where $Z_{{\rm AdS}_3}$ is the partition function of the AdS$_3$ gravity with the scalar fields $\phi$ of mass $m^2=\Delta(\Delta-2)$ with the Dirichlet boundary condition as indicated. One can then rest assured that the gravity dual reproduces the same results as those of the field theory. 


\section{Discussions}
\label{Sec:Discussions}

We studied Cardy's random geometry approach to the $T\bar{T}$ deformation of 2$d$ CFT \cite{Cardy:2018sdv} with the aim of finding a gravity dual of $T\bar{T}$-deformed CFTs.
In this representation of the $T\bar{T}$ deformation, the gravity dual description becomes a straightforward translation of languages from field theory to gravity.
As a result, the gravity dual is an ensemble of AdS$_3$ spaces or BTZ black holes with randomly fluctuating boundary metrics over the scale of $T\bar{T}$ deformation.

\medskip
In the course of our discussions, we streamlined the method of computation in the random geometry approach and provided new simplified ways of deriving the energy spectrum \cite{Smirnov:2016lqw, Cavaglia:2016oda} and the thermal free energy \cite{Cardy:2018sdv}. We further generalized this approach to the computation of correlation functions and showed how the results in \cite{Cardy:2019qao} can be reproduced in this new method. 
As further applications of our new method, both from the field theory and gravity dual viewpoints, albeit obvious ideas, interesting things to do is to generalize the analysis of Section \ref{Sec:EnergySpectrum} to the case with momentum flow, corresponding in the gravity dual to rotating BTZ black holes, and the computation of Section \ref{Sec:Correlators} to stress tensor correlators \cite{Aharony:2018vux} and correlators on other topologies.

\medskip
Related technically to the last point is a study of chaos in $T\bar{T}$-deformed CFTs and their gravity duals by analyzing out-of-time-order correlators (OTOC). The undeformed CFTs saturate the chaos bound \cite{Maldacena:2015waa} and it is of interest to understand what correction to Lyapunov exponents the $T\bar{T}$ deformation yields. Since the coupling $\mu$ has length dimension 2, similar to $\alpha'$ of string theory, and the energy spectrum is of the Nambu-Goto type \cite{Dubovsky:2012wk, Caselle:2013dra}, it might be that the correction is similar to a stringy effect discussed in \cite{Shenker:2014cwa}.

\medskip
One of the most peculiar features of $T\bar{T}$-deformed theories is the appearance of complex energy for $\mu>0$ and the Hagedorn-like phase for $\mu<0$.
A wild speculation is that since a complex energy may be thought of as an instability or a loss of energy, the fluctuating boundary metrics effectively act as an absorbing boundary condition for $\mu>0$ such that high energy quanta above a certain threshold are lost into the boundary. In contrast, for $\mu<0$, the boundary may become emitting and the energy of the system might reach infinity before the temperature becomes infinitely high.
As another speculation, pushing an analogy between $\mu$ and $\alpha'$, it might be that a positive/negative $\mu$ corresponds to a negative/positive tension of strings. In this scenario, a negative $\mu$ theory behaves much like string theory and exhibits a Hagedorn behavior, whereas a positive $\mu$ theory becomes unstable. We hope that our gravity dual description can help us better understand these puzzling features of $T\bar{T}$-deformed theories.

\medskip
We expect that these unconventional features will be manifested in entanglement and Renyi entropies. Namely, for shorter intervals, entanglement and Renyi entropies will presumably become complex for $\mu>0$, whereas they will vanish at some finite minimal interval for $\mu<0$.
This may be studied most conveniently by the Ryu-Takayanagi prescription \cite{Ryu:2006bv}. 
(For earlier works on entanglement and Renyi entropies in $T\bar{T}$-deformed CFTs, see, for example, \cite{Donnelly:2018bef, Chen:2018eqk, Caputa:2019pam} which are based on the cutoff AdS proposal of \cite{McGough:2016lol}.)

\medskip
It is natural to ask whether our gravity dual can, in some way, be
related to two other proposals, i.e.~the cutoff AdS by
McGough-Mezei-Verlinde \cite{McGough:2016lol} and the mixed nonlinear
boundary condition by Guica-Monten \cite{Guica:2019nzm}. However, we
first note that neither of these two proposals can successfully
reproduce matter correlation functions. Moreover, even though the energy
spectrum is reproduced in a nontrivial and remarkable manner in their
proposals, the way how it matches with the field theory result differs
from ours.  At least, it seems clear that in the presence of local
operators on the boundary, the cutoff surface cannot be
a constant $r$ surface.
Nevertheless, we feel that with some refinements, these two
proposals will work and can be related to our version of the gravity
dual. A possible connection to the cutoff AdS proposal might be made via
holographic Wilsonian renormalization group (RG) suggested in
\cite{Heemskerk:2010hk} if the Gaussian $h$ action of the
Hubbard-Stratonovich transformation arises by integrating out (field
theory) UV degrees of freedom in a similar way to how double-trace
operators are induced under a holographic Wilsonian RG flow.

\medskip
The generalization of $T\bar{T}$ deformations to higher dimensions was proposed in \cite{Taylor:2018xcy, Hartman:2018tkw}. The 3$d$ $TT$-deformed CFTs may have an interesting application to inflation cosmology via the dS/CFT correspondence as a computational tool or an effective theory \cite{Maldacena:2002vr} because the flow to UV corresponds to the forward time evolution of an inflating universe \cite{Strominger:2001gp} and a weak $TT$-deformation can be regarded as a slow-roll. In \cite{Cardy:2018sdv} an attempt was made to generalize the random geometry approach to higher dimensions, in particular, to three dimensions. 
Although life is not quite as simple as in two dimensions, if a similar method can be developed for correlation functions in the 3$d$ case, it may provide a model-independent way of calculating the power, bi- and tri-spectra in the cosmic microwave background.

\medskip
As yet another application, it would be interesting to study the effect of the $T\bar{T}$ deformation on drag force in AdS/CFT \cite{Gubser:2006bz}. 
Drag force is a characteristic phenomenon that occurs in AdS black holes: an open string, dual to a heavy quark, moving in thermal radiation of black holes, dual to a quark-gluon plasma (QGP), experiences a drag from the medium. In the case of BTZ black holes, it was studied in \cite{Herzog:2006gh} and, more recently, further generalized  in \cite{Bena:2019wcn} to horizonless microstate geometries \cite{Iosif, David:2002wn}.
Based on an observation made in \cite{Cardy:2015xaa}, we expect that the effect of the $T\bar{T}$ deformation is a renormalization of propagation speed in the speed dependence of the drag force.

\medskip
Finally, it would be interesting to apply our method of $T\bar{T}$-deformed correlator computation to the Liouville theory since it would correspond to a new integrable deformation of ${\cal N}=2$ supersymmetric $SU(2)$ gauge theories \cite{Seiberg:1994rs} that is yet to be uncovered via the AGT correspondence \cite{Alday:2009aq}.


\section*{Acknowledgments}

We would like to thank Ofer Aharony and Sam van Leuven for discussions.
%
%
SH would like to thank Nagoya University and Technion for their
hospitality where part of this work was done.  The work of SH was
supported in part by the National Research Foundation of South Africa
and DST-NRF Centre of Excellence in Mathematical and Statistical
Sciences (CoE-MaSS).  Opinions expressed and conclusions arrived at are
those of the authors and are not necessarily to be attributed to the NRF
or the CoE-MaSS.  The work of MS was supported in part by JSPS KAKENHI
Grant Numbers 16H03979, and MEXT KAKENHI Grant Numbers 17H06357 and
17H06359.

\appendix

\section{ADM mass of BTZ black holes}
\label{Appendix:ADMmass}

The BTZ black holes \cite{Banados:1992wn} in the Fefferman-Graham coordinates are described by
\begin{align}
ds_{\rm BTZ}^2=-N^2dt^2+g_{yy}\left(dy-{Jdt\over 2g_{yy}}\right)^2+{d\rho^2\over\rho^2}\ ,
\end{align}
where $y\sim y+2\pi R$ and 
\begin{align}
\hspace{-.3cm}
N&=\sqrt{-g_{tt}+{J^2\over 4g_{yy}}},\quad g_{yy}=\frac{\rho ^2}{4}+\frac{M}{2}+\frac{M^2-J^2}{4 \rho ^2},\quad
g_{tt}=-\left(\frac{\rho ^2}{4}-\frac{M}{2}+\frac{M^2-J^2}{4 \rho ^2}\right).
\end{align}
The conserved (ADM) mass is defined by \cite{Brown:1992br} 
\begin{align}
M_{\rm ADM}=\int_{S^1}dy\sqrt{g_{yy}}u^tT_{tt}\xi^t=\int_{S^1}dy\sqrt{g_{yy}}N(u^tT_{tt}u^t)
\end{align}
where $u^t$ is a unit timelike tangent vector along the boundary and $\xi^t=Nu^t$ is a timelike Killing vector with the lapse $N$. The Brown-York tensor is defined by
\be
T_{ab}={1\over 8\pi G}\left(\Theta_{ab}-h_{ab}\Theta\right)\qquad\quad\mbox{with}\qquad\quad \Theta_{ab}=-\half\left(\nabla_an_b+\nabla_bn_a\right)\ .
\ee
The unit normal vector is 
\be
n_a=\left(0, 0, \rho^{-1}\right)\qquad\quad\mbox{with}\qquad\quad g^{ab}n_an_b=1
\ee
and the unit timelike tangent vector is
\be
u^t=\left(N^{-1},0,0\right)\qquad\quad\mbox{with}\qquad\quad g^{ab}t_at_b=-1\ .
\ee
The extrinsic curvature can be calculated as
\begin{align}
\Theta_{\mu\nu}=\left(
\begin{array}{ccc}
\frac{J^2-M^2+\rho ^4}{4 \rho ^2} & 0 & 0 \\
0 & -\frac{J^2-M^2+\rho ^4}{4 \rho ^2} & 0 \\
0 & 0 & 0
\end{array}
\right)\ .
\end{align}
Thus the Brown-York tensor can be found as
\begin{align}
T_{ab}=\left(
\begin{array}{cc}
-\frac{J^4-2 J^2 \left(M^2-2 M \rho ^2+3 \rho ^4\right)+\left(M-\rho ^2\right)^4}{4 \rho ^2 \left(J^2-M^2+\rho ^4\right)} 
& \frac{J \left(J^2-M^2-\rho ^4\right)}{J^2-M^2+\rho ^4}  \\[1ex]
\frac{J\left(J^2-M^2-\rho ^4\right)}{J^2-M^2+\rho ^4} 
& \frac{J^4-2 J^2 \left(M^2+2 M \rho ^2+3 \rho ^4\right)+\left(M+\rho ^2\right)^4}{4 \rho ^2 \left(J^2-M^2+\rho ^4\right)}
\end{array}
\right)\ .
\end{align}
Taking into account the rescaling
\be\label{App:rescaling}
t\to \sqrt{1+h_{11}}\,t\ ,\qquad 
y\to \sqrt{1+h_{22}}\,y\ ,
\ee
we find the (bare) ADM mass  in the unit $8G=1$
\begin{align}
M^{(bare)}_{\rm ADM}=2\pi R\sqrt{g_{yy}}N^{-1}T_{tt}=R\sqrt{(1+h_{11})(1+h_{22})}\left(M-\half\rho^2\right)+{\cal O}(1/\rho^2)
\end{align}
which can be holographically renormalized as \cite{Balasubramanian:1999re, Emparan:1999pm, deHaro:2000vlm}
\begin{align}\label{ADMmassren}
M_{\rm ADM}=2\pi R\sqrt{g_{yy}}N^{-1}\biggl(T_{tt}-{1\over 8\pi G}g_{tt}\biggr)=R\sqrt{(1+h_{11})(1+h_{22})}M+{\cal O}(1/\rho^2)\ .
\end{align}


\section{On-shell action of Euclidean BTZ black holes}
\label{Appendix:Onshellaction}

The (Euclidean) gravity action is given by
\begin{align}\label{gravityaction}
-S_E={1\over 16\pi G}\int_{\cal M} d^{d+1}x \sqrt{g}\left(R+{d(d-1)\over\ell^2}\right)-{1\over 8\pi G}\int_{\del{\cal M}} d^d x\sqrt{\gamma}\Theta
-{1\over 8\pi G}\int_{\del{\cal M}} d^d x\sqrt{\gamma}
\end{align}
with $d=2$ and $\ell=1$. The second term is the Gibbons-Hawking-York boundary term and the third is the holographic counter-term.
Taking into account the rescaling \eqref{App:rescaling} that amounts to a resizing of the volume $A$, the on-shell bulk action reads
\begin{align}
S_{\rm bulk}={A\over 16\pi G}\sqrt{(1+h_{11})(1+h_{22})}{1\over 2\rho^2}\left[\rho^4-(4GJ)^2+(8GM)^2\right]^{\rho=\Lambda}_{\rho=\rho_h}
\end{align}
where $\rho_h=\sqrt{4GJ+8GM}$ is the location of the horizon. 
Next, the GHY term and the counter-term are given, respectively, by
 \begin{align}
S_{\rm GHY}=-{A\over 16\pi G}\sqrt{(1+h_{11})(1+h_{22})}\Lambda^2\ ,\quad
S_{\rm ct}={A\over 16\pi G}\sqrt{(1+h_{11})(1+h_{22})}\half\Lambda^2\ .
\end{align}
Hence the renormalized on-shell action is found to be
\begin{align}\label{onshellwocone}
S_{\rm EBTZ}=S_{\rm bulk}+S_{\rm GHY}+S_{\rm ct}=-L RM\sqrt{(1+h_{11})(1+h_{22})}
\end{align}
where we used $A=2\pi RL$.


\end{document}